\documentstyle[preprint,tighten,aps]{revtex}

\begin{document}
\author{Yu.L.Klimontovich}
\title{To Phenomenological Theory of Superconductivity. Superconductivity - not
fading electrical current in dissipative medium}
\maketitle

\begin{abstract}
The basic stages of development of the theory of superconductivity are
traced. Despite of remarkable successes of theory, the physical explanation
of the phenomenon of superconductivity - of the not fading electrical
current in dissipative medium - is not exists until now.

In the present paper on the basis of works (Klimontovich, 1990abc, 1999) the
physical explanations of this phenomenon is considered. Will be show that
the existence of not fading electrical current becomes possible due to
occurrence of flicker noise and appropriate residual temporary correlations.
\end{abstract}

\section{Introduction}

In 1911 Dutch physicists H.Kamerlingh-Onnes has found out at temperature
4.12 the superconductivity of mercury - the electrical resistance of mercury
suddenly disappeared and remained equal the zero at the further downturn of
temperature. A bit later superconductivity was found out and in others
metals and also in some alloys.

Long years was supposed, that the superconductivity is the phenomenon of low
temperature physics. The high-temperature superconductivity was found out
for the first time only in 1986 by Bednorz and Muller.

In 1933 by Meissner and Ochxenfeld have found out that the weak magnetic
field is pushed out from a massive superconductor - ''Meissner effect''. The
value of a superconducting current in a sample, which is a part of a
consecutive circuit, by a source of a current in the circuit is determined.
The superconductivity disappears, when the value of a current reaches some
critical value.

The first phenomenological theory of superconductivity was proposed in
brothers London work in 1935 (see. London, 1954). They have admitted the
existence of the superconductivity. On this basis they have shown, that the
Meissner effect is inevitable consequence of the superconductivity
phenomenon. A question on validity of the opposite statement remained open.

In 1950 on the basis of London equations the phenomenon of quantization of
magnetic flow was predicted. It was revealed almost simultaneously by two
groups of experimenters in 1961. They showed that the quant of magnetic flow
is defined by the particles with the double of electron charge. This result
has confirmed the Cooper effect. The last served as the basis of the Bardeen
- Cooper - Schriffer (BCS) microscopic theory of superconductivity (1957).

In 1950 the Ginsburg - Landau (GL) the stationary equation for some
effective wave function have offered. The coefficients of GL equation by
measured values of the critical magnetic field and the by the London
penetration depth are defined.

The appreciable contribution to the theory of superconductivity was made By
Bogolubov (1958). In the Gor'kov paper (1959) the connection of GL and BCS
theories was established.

After opening of high-temperature superconductivity new mechanisms of this
phenomenon were discussed. However the enough convincing microscopic theory
of the high-temperature theory in the present time yet does not exists (see
Abrikosov, 1987; Plakida, 1996). In such situation the phenomenological
description of superconductivity plays the important role.

The of London and GL equations are not dissipative and correspond to
approximation of continuous medium. As all equations in approximation of
continuous medium are generally dissipative, that the following basic
question arises: How in the dissipative system exists not fading electrical
current?

The possible answer on this question was discussed earlier (Klimontovich,
1990abc, 1995). It was shown that the existence of not fading electric
current becomes possible due to occurrence of flicker noise and appropriate
residual temporary correlations (Kogan, 1985; Klimontovich, 1982, 1983). In
present paper this works serve by the basis of more general description of
the superconductivity.

The possibility of connection between the two seemingly antagonistic
phenomena: flicker noise and superconductivity has the following explanation.

First, both, flicker noise and superconductivity are spatial coherent
phenomena. In the domain of flicker noise the distribution on wave numbers
has a very sharp maximum near to its zero value. Moreover, the dispersion is
proportional to frequency, and therefore tends to zero together with $\omega 
$. Thus has a place original Bose condensation

\section{The GL and London equations}

Let's designate number of Cooper pairs through $n^{\ast }(T)$ and shall
enter the effective wave function of pairs of superconductive electrons $%
\psi (R,t)$.

The Debye radius is much smaller of characteristic scale of Cooper pairs,
together with size of a point of continuous medium. On this reason the
Coulomb interaction not play an essential role in processes of
superconductivity. The square of the module of function $\psi (R,t)$ defines
average density of number of electron pairs 
\begin{equation}
\left| \psi (R,t)\right| ^{2}\equiv n^{\ast }(T)=\frac{n_{S}(T)}{2}.
\label{S.2.1}
\end{equation}
In the GL work the stationary equation for the effective wave function is
considered only. The one from opportunities of the temporary description -
the use for of effective wave function, appropriate, the nonlinear
Schroedinger equation (Hartree equation) for particles with the double
electron charge $e^{\ast }=2e$. Just so acts, for example, Feinman, (1967. 
\begin{equation}
I\hbar \frac{\partial \psi (R,t)}{\partial t}=\frac{1}{2m^{\ast }}\left|
\left( -i\hbar \frac{\partial }{\partial R}-\frac{e^{\ast }}{with}A\right)
\right| ^{2}\psi (R,t)+\varphi (R,t)\psi ,\quad divA=0.  \label{S.2.4}
\end{equation}
The GL potential $\varphi _{GL}(R,t)$ will be defined below.

Let's wrote the appropriate equation of a continuity

{\sl 
\begin{equation}
\frac{\partial \left| \psi (R,t\right| ^{2}}{\partial t}+\frac{\partial
j_{S}(R,t)}{\partial R}=0.  \label{S.2.4a}
\end{equation}
} The current created by electron pairs is defined by expression: 
\begin{equation}
j_{S}(R,t)=-\frac{ie^{\ast }\hbar }{2m^{\ast }}\left( \psi ^{\ast }\nabla
\psi -\psi \nabla \psi ^{\ast }\right) -\frac{e^{\ast 2}}{2m^{\ast }c}\left|
\psi \right| ^{2}A(R,).  \label{S.2.5}
\end{equation}
Let's present complex function $\psi (R,t)$ as

\begin{equation}
\psi (R,t)=\left| \psi (R,t)\right| \exp \left( i\theta (R,t)\right) \equiv
a(R,t)\exp \left( i\theta (R,t)\right) {\rm .}  \label{S.2.6}
\end{equation}
Then the electric current created by the superconductive pairs is defined by
expression:

\begin{equation}
j_{S}(R,t)\equiv e^{\ast }\left| \psi (R,t\right| ^{2}u_{S}(R,t)=\frac{%
e^{\ast }\hbar }{m^{\ast }}\left( \frac{\partial \theta }{\partial R}-\frac{%
e^{\ast }}{\hbar c}A\right) \left| \psi (R,t\right| ^{2}.  \label{S.2.7}
\end{equation}
From here the expression for the mean velocity of superconductive electron
follow:
\begin{equation}
u_{S}(R,t)=\frac{\hbar }{m^{\ast }}\frac{\partial \theta }{\partial R}{}-%
\frac{e^{\ast }}{m^{\ast }c}A,\quad divA=0.  \label{S.2.8}
\end{equation}

It consists from the sum of the potential and the vortical components. The
current $j_{S}(R,t)$ in a normal state is equal to zero.

In massive metals the electron density $\left| \psi (R,t\right| ^{2}=const.$
and the Maxwell equations look like 
\begin{equation}
rotB=\frac{4\pi }{c}j_{S}^{,}\quad E=0.\ B=rotA,\quad divj_{S}=0.
\label{S.2.9}
\end{equation}
Under condition of a constancy of a phase $\theta (R)=const,$, that has a
place for continuous (without holes) superconductor, the potential component
of the velocity $u_{S}(R,t)$ is equal to zero and the current is connected
to the vector potential by the equation of London
\begin{equation}
j_{S}(R)=-\ \left| \psi (R,t\right| ^{2}\frac{e^{\ast 2}}{m^{\ast }c}A\equiv
-\ \left| \psi (R,t\right| ^{2}\frac{2e^{2}}{mc}A.  \label{S.2.12}
\end{equation}

From it in a combination to the Maxwell equations the closed equation for
the magnetic field follows:

\begin{equation}
\frac{\partial ^{2}B}{\partial R^{2}}+\frac{1}{\delta _{L}^{2}}B=0.
\label{S.2.14}
\end{equation}
Designation for the depths of penetration - the London parameter $\delta _{L}
$here is used

\begin{equation}
\delta _{L}^{2}=\frac{m^{\ast }c^{2}}{4\pi e^{\ast 2}\left| \psi (T)\right|
^{2}}\equiv \frac{mc^{2}}{4\pi e^{2}n_{S}},  \label{S.2.15}
\end{equation}
Thus the theory of London, in which the fact of the existence, of not fading
electrical current is accepted, gives the explanation of Meissner effect.
The expression, received by such way, for penetration depth $\delta _{L}$,
corresponds to experimental data. Typical size $\delta _{L}$ of the order 10 
$^{^{-5}}$ cm.

So, the theory of London is based on the assumption of the existence of a
not fading electrical current, and gives the explanation of Meissner effect.
The equations describing the screening of magnetic field and the current are
classical ones - does not contain the Planck constant.

At an establishment of the equation of London the equation of continuity was
used only . Let's show, that the equation of London satisfies to complete
system of the reversible hydrodynamical equations, which turns out on the
basis of the Schroedinger equation.

Being based on the Schroedinger equation for the effective wave function it
is possible to receive and equation for average velocity of superconducting
electrons:

\begin{equation}
\frac{\partial u_{S}}{\partial t}+\left( u_{S}\frac{\partial }{\partial R}%
\right) u_{S}=\frac{e^{\ast }}{m_{\ast }}\left( E+\frac{1}{c}\left[ u_{S}B%
\right] \right) +\frac{1}{m_{S}}\frac{\partial U_{quant}}{\partial R}.
\label{S.2.31}
\end{equation}
Designation is entered here 
\begin{equation}
U_{quant}=\frac{\hbar ^{2}}{2\sqrt{\rho _{S}}}\frac{\partial ^{2}\sqrt{\rho
_{S}}}{\partial R^{2}}  \label{S.2.32}
\end{equation}
for " the quantum potential energy", through which enters a quantum source".

In the theory of London a magnetic field and, as a consequence, and
hydrodynamical velocity are small. In linear approximation the equation of
motion has form:

\begin{equation}
\frac{\partial u_{S}}{\partial t}=-\frac{e^{\ast }}{m^{\ast }\tilde{n}}\frac{%
\partial A}{\partial t}.  \label{S.2.33}
\end{equation}
This equation is consequence of the Schroedinger equation, therefore does
not contain a dissipation. In view of a constancy of the density of electron
pairs can it be rewrite as the London equation

\begin{equation}
u_{S}(R)=-\frac{e^{\ast }}{m^{\ast }c}A(R).  \label{S.2.34}
\end{equation}
Thus the reversible equation (\ref{S.2.33}) is carried out just by virtue of
the London equation.

The London theory does not explain the phenomenon of existence of
superconductive current, but essentially bases on fact of its existence. The
condition is used also: $\left| \psi (R,t\right| ^{2}=const$ is used also.
Thus the question on dependence of number of superconducting electrons on
temperature there is without the answer.

\section{Comparison with the GL theory}

The following essential step in development of the phenomenological theory
of superconductivity was made in work GL. However, the nature of phase
transition in superconductors was not concretized. It was found out only at
creation if the BCS theory.

The GL theory not give also the answer on the basic question: Why in
superconductor, which, as any continuous medium is a dissipative system, is
possible not fading electrical current?

The stationary GL equation has a form: 
\begin{equation}
-\frac{\hbar ^{2}}{2m^{\ast }}\frac{\partial ^{2}}{\partial R^{2}}+\left(
\alpha \frac{T-T_{C}}{T_{C}}{}{}{}{}{}+b\left| \psi \right| ^{2}\right) \psi
=0.  \label{2.1.2}
\end{equation}

It presents the "synthesis" of the quantum mechanics and the Landau theory
of a second order phase transition. The BCS theory allows to connect
coefficients $\alpha ,b$ of the GL theory with parameters of a normal state
of superconductor.

There are two opportunities of temporary generalization of this equation.
Above was is used the quantumechanic generalization. It has resulted to the
nonlinear Schroedinger equation (\ref{S.2.4}). It corresponds to the Hartree
equation with the GL potential
\begin{equation}
\varphi _{G-L}=\alpha \frac{T-T_{C}}{T_{C}}{}{}{}{}{}+b\left| \psi \right|
^{2}.  \label{2.1.2a}
\end{equation}

Above at the description of temporary evolution the preference to quantum
was given up. In result came to the nonlinear Schroedinger equation (the
Hartree equation):

\begin{equation}
i\hbar \frac{\partial \psi }{\partial t}=-\frac{1}{2m^{\ast }}\left| \left(
-i\hbar \frac{\partial }{\partial R}-\frac{2e}{c}A\right) \right| ^{2}\psi
+\varphi _{G-L}\psi .  \label{2.5.1}
\end{equation}
The members responsible for phase transition, are expressed through
appropriate potential - potential ''GL'':

\begin{equation}
\varphi _{GL}=\left( \alpha \frac{T-T_{C}}{T_{C}}+b\left| \psi \right|
^{2}\right) .  \label{2.5.2}
\end{equation}
The presence of additional potential $\varphi _{GL}$ does not change a form
of the continuity equation. The second hydrodynamical equation accepts the
form:

\[
\frac{\partial u_{S}}{\partial t}+\left( u_{S}\frac{\partial }{\partial R}%
\right) u_{S}=\frac{e^{\ast }}{m^{\ast }}\left( E+\frac{1}{c}\left[ u_{S}B%
\right] \right) - 
\]
\begin{equation}
-\frac{1}{m^{\ast }}\frac{\partial \left( U_{quant}+\varphi _{G-L}\right) }{%
\partial R}.  \label{2.5.3}
\end{equation}
Through the additional force the dependence of number of superconductive
electron pairs from temperatures enters. This dependence in the London
theory is not taken into account. Under former conditions (constancy of a
phase and density of superconducting electrons) the London equation
satisfies in linear approximation to he last equation.

The Hartree equation with potential GL is reversible and, gives only dynamic
description of a superconductor. As, however, the number the superconducting
electrons pairs depends on temperature, that more adequate is not "dynamic",
but ''chemical'' way of temporary generalization of the stationary GL
equation. This question repeatedly was discussed in the literature (see, for
example, Elesin and Kopaev, 1981; Oraevskii, 1993). Let's consider the
appropriate generalization of the G-L equation for the description of
temporary processes in superconductors.

\section{Relaxation GL and kinetic equations}

Let's give back now the preference to dissipative processes, which take
place at phase transitions. Thus instead of the GL equation we come to the
relaxation GL equation (RGLE).

For system of superconducting pairs the characteristic length - the
coherence length at zero temperature and the coefficient of spatial
diffusion are defined by expressions:

\begin{equation}
\xi _{0}^{2}=\frac{\hbar ^{2}}{2m^{\ast }\alpha }\approx \frac{p_{F}^{2}}{%
p_{T_{C}}}\lambda _{B}^{2};\quad D=\frac{\hbar }{2m^{\ast }}.  \label{2.6.1}
\end{equation}
Expressions for time diffusion time and, appropriate, from here follow of
the friction coefficient from here follow

\begin{equation}
\tau _{D}=\frac{1}{\gamma }=\frac{\xi _{0}^{2}}{D}=\frac{\hbar }{\alpha }%
;\quad \gamma =\frac{\alpha }{\hbar }.  \label{2.6.2}
\end{equation}
The RGLE at ($A=0$) has form:

\begin{equation}
\frac{\partial \psi (R,t)}{\partial t}=-\text{ }\frac{1}{2}\gamma \left( 
\frac{T-T_{C}}{T_{C}}+\frac{\left| \psi \right| ^{2}(R,t)}{n}\right) \psi +D%
\frac{\partial ^{2}\psi (R,t)}{\partial R^{2}}.  \label{2.6.3}
\end{equation}
It serves an example of the reaction diffusion equation. In it interactions
with normal electrons and phonons, dissociation and formation of Cooper
pairs it are taken into account only through coefficients.

The RGLE follows from the Schroedinger equation by formal replacement of the
time $t$ by imaginary time $it$. Let's consider other way of the description
of relaxation processes, which is, indubitably, more consecutive.

Let's begin from a case, when the distribution of values of phase is not
essential. The appropriate Fokker-Planck equation for the local distribution
function $f(n^{\ast },R,t)$ of values of $\left| \psi \right| ^{2}=n^{\ast }$
has the following form:

\begin{equation}
\frac{\partial f}{\partial t}=\frac{\partial }{\partial n^{\ast }/n}\left[
D_{n^{\ast }}\frac{n^{\ast }}{n}\frac{\partial f}{\partial n^{\ast }/n}%
\right] +\frac{\partial }{\partial n^{\ast }/n}\left[ \gamma \left( \frac{%
T-T_{C}}{T_{C}}+\frac{n^{\ast }}{n}\right) \frac{n^{\ast }}{n}f\right] +D%
\frac{\partial ^{2}f}{\partial R^{2}}.  \label{2.6.4}
\end{equation}
The diffusion coefficient $D_{n^{\ast }}$ is defined by the expression:

\begin{equation}
\qquad D_{n^{\ast }}=\frac{1}{N_{ph}}\frac{k_{B}T}{\hbar }.  \label{2.6.5}
\end{equation}
This results to appropriate the smoothed Boltzmann distribution.

In the self-consistent approximation on the first moment (when $f(n^{\ast
},R,t)=\delta (n^{\ast }-n^{\ast }(R,t)$) we have the following equation:

\begin{equation}
\frac{\partial n^{\ast }(R,t)/n}{\partial t}=\left[ D_{n^{\ast }}-\gamma
\left( \frac{T-T_{C}}{T_{C}}+\frac{n^{\ast }(R,t)}{n}\right) \frac{n^{\ast
}(R,t)}{n}\right] +D\frac{\partial ^{2}n^{\ast }(R,t)/n}{\partial R^{2}}.
\label{2.6.7}
\end{equation}
In it two important factors are not yet taken into account: evolution of a
phase and action electrical and magnetic fields. Without its the explanation
neither superconductivity, nor Meissner effect it is impossible. The
corresponding more general kinetic equation will be carried out below.

In the stationary and spatial - homogeneous state is received the algebraic
equation for function $n^{\ast }(T)$:
\begin{equation}
\left( \frac{n^{\ast }(R,t)}{n}\right) ^{2}+\frac{T-T_{C}}{T_{C}}\frac{%
n^{\ast }(R,t)}{n}=\frac{D_{n^{\ast }}}{\gamma }=\frac{1}{N_{ph}}\frac{k_{B}T%
}{\hbar \gamma }=\frac{1}{N_{ph}}\frac{k_{B}T}{\alpha }.  \label{2.6.8}
\end{equation}

This equation defines a finite solution at all values of temperature. For
temperatures is significant larger critical one, the solution has following
form:

\begin{equation}
n^{\ast }(R,t)=\frac{1}{N_{ph}}\frac{k_{B}T}{\alpha }\frac{T_{C}}{T-T_{C}}n;
\label{2.6.10}
\end{equation}
In the critical point

\begin{equation}
n^{\ast }(R,t)=\sqrt{\frac{1}{N_{ph}}\frac{k_{B}T}{\alpha }}n  \label{2.6.11}
\end{equation}
and, at last, for temperatures, below critical, the density of numbers of
superconducting electrons is defined by expression: 
\begin{equation}
n^{\ast }(R,t)=\frac{T_{?}-T}{T_{C}}n,  \label{2.6.12}
\end{equation}
which coincides with result of the Landau theory.

\section{Whether exists now theory of superconductivity?}

Two variants of temporary generalization of the stationary GL were
considered. In the first case temporary evolution is described on the basis
of the reversible Hartree equation with potential, which is determined by
distribution of density of superconducting electrons. As well as in the
London theory, still have open a question on a nature of superconductive
electric current in a dissipative medium. In the second case were used the
appropriate relaxation equation.

In both cases there is open the question on a physical nature not fading
electrical current. Thus despite of doubtless successes, rather complete
theory the superconductivity now does not exist yet.

In such situation the aspiration is natural to construct the evolutionary
equations for the description of superconductivity with the simultaneous
account as dynamic, so and the dissipative contributions.

Will be shown, that the existence of the superconducting current becomes
possible due to occurrence of flicker noise and, appropriate, residual
temporary correlations.

\section{Dynamic reaction diffusion equation in the theory superconductivity}

Was shown that hydrodynamical equations at rather small velocity does not
contain the Planck constant. Let's remind also, that classical expression
for the London length invariant concerning replacement:

\begin{equation}
e,m,n,\longleftrightarrow e^{\ast },m^{\ast },n^{\ast }.  \label{12.12.7}
\end{equation}
and in the theory of London there is no dependence on a kind of statistics:
Fermi or Bose. The role of the electron pairs is reduced to the following.

The first, due to them in the nonideal fermi-gas there is a phase transition
- there is the equilibrium state with lower energy, than appropriate state
of free electrons.

The secondly, the superconductive state represents continuous medium, in
which the size of a point is defined by the quantum parameter - the de
Broglie wave length $\lambda _{B}=\hbar /p_{T_{C}}.$ This length
considerably exceeds the Debye radius. Thanks to this, the superconductor
represents unusual continuous quantum medium. In it " physical Knudsen
number" - the basic small parameter of continuous medium, is defined by
expression:

\begin{equation}
\left( K_{n}\right) _{ph}=\frac{\lambda _{B}}{L}=\frac{\hbar }{p_{T_{C}}L}
\label{12.12.8}
\end{equation}
and, hence, is the quantum characteristic. For a superconductor role of
parameter $L$ play the size of the Cooper pair $\xi _{0}=\hbar /\sqrt{%
2m^{\ast }\alpha }$ and the London length $\lambda _{B}$. For
superconductors of the London type the greatest is dimensionless small
parameter (Klimontovich, 1995):

\begin{equation}
\left( K_{n}\right) _{ph}=\frac{\lambda _{B}}{\xi _{0}}\approx \frac{%
p_{T_{C}}}{p_{F}}\ll 1.  \label{12.12.9}
\end{equation}
Thus, the phase transition in the superconducting state, caused by existence
of the Cooper pairs, it is possible to treat as transition to quantum
continuous medium. By the basic dimensionless small parameter which serves
the quantum physical Knudsen number.

Be taking into account that the dynamic in the London theory is classical,
the dissipative kinetic equation for the local distribution function $%
f(n^{\ast },R,v,t)$ of values of density of electron pairs it is possible to
write in form:

With the account of dissipation for the equation for the local distribution
function $f(n^{\ast },R,v,t)$ of values of density of pairs of
superconducting electrons it is possible to write down as:

\begin{equation}
\frac{\partial f}{\partial t}+v\frac{\partial f}{\partial R}{}+\frac{e^{\ast
}}{m^{\ast }}\left( E(R,t)+\frac{1}{c}\left[ u_{S}B(R,t)\right] \right) 
\frac{\partial f}{\partial v}{}=I_{(v)}+I_{(R)}+I_{(n^{\ast })};\quad divE=0
\label{12.12.10}
\end{equation}
The norm of the distribution function $f(n^{\ast },R,v,t)$ will be carried
out on the average of particles, which depends on temperature.

The sum of second and third "collision integrals" we shall present as the
sum of two contributions. First describes spatial diffusion of distribution
function. Second - diffusion in the space of values of number of electron
pairs:

\begin{equation}
I_{(R)}+I_{(n^{\ast })}=D\frac{\partial ^{2}f}{\partial R^{2}}+\frac{%
\partial }{\partial n^{\ast }}\left[ D_{n^{\ast }}nn^{\ast }\frac{\partial f%
}{\partial n^{\ast }}\right] +\frac{\partial }{\partial n^{\ast }}\left[
\gamma \left( \frac{T-T_{C}}{T_{C}}+\frac{n^{\ast }}{n}\right) n^{\ast }f%
\right] .  \label{12.12.12}
\end{equation}
The diffusion $D_{n^{\ast }},$ $D$ and friction $\gamma $ coefficients are
defined by the formulas:
\begin{equation}
D_{n^{\ast }}=\frac{1}{N_{ph}}\frac{k_{B}T}{\hbar };\quad D=\hbar /2m^{\ast
};\quad \gamma =\frac{\alpha }{\hbar }.  \label{12.12.13}
\end{equation}
''The collision integral'' $I_{(v)}$ defines redistribution of pairs of
electrons on velocity. It has the same properties, as the Boltzmann
collision integral in the kinetic theory of gases.

Let's proceed from the kinetic equation to the equations for local functions 
$\left\langle n^{\ast }\right\rangle _{R,t},$ $u_{S}(R,t).$ \ In The
self-consistent approximation on the first moments for the distribution
function $f(n^{\ast },R,v,t)$ it is represented as:

\begin{equation}
f(n^{\ast },R,v,t)=\delta \left( v-u_{S}(R,t)\right) \delta \left( n^{\ast
}-\left\langle n^{\ast }\right\rangle _{R,t}\right) .  \label{12.12.19}
\end{equation}
In result is received the equation for average local density number
particles:

\begin{equation}
\frac{\partial }{\partial t}\frac{\left\langle n^{\ast }\right\rangle _{R,t}%
}{n}+\frac{\partial }{\partial R}\frac{\left\langle n^{\ast }\right\rangle
_{R,t}u_{S}(R,t)}{n}=D_{n_{S}}-\gamma \left( \frac{T-T_{C}}{T_{C}}+\frac{%
\left\langle n^{\ast }\right\rangle _{R,t}}{n}\right) \frac{\left\langle
n^{\ast }\right\rangle _{R,t}}{n}+D\frac{\partial ^{2}}{\partial R^{2}}\frac{%
\left\langle n^{\ast }\right\rangle _{R,t}}{n}.  \label{12.12.20}
\end{equation}
It contains the members, taking into account birth and disappearance of
electron pairs ("chemical reaction") and a self-diffusion contribution. To
this reason the flow of electron pairs is defined not only the convective
transfer $\left\langle n^{\ast }\right\rangle _{R,t}u_{S}(R,t)$, but also by
the spatial diffusion. The similar results take place at the description of
nonequilibrum processes in gases and plasma (Klimontovich, 1995.1999).

In the London equations the electron pairs density $\left\langle n^{\ast
}\right\rangle _{R,t}=const.$ Now in the stationary state we have the
equations:

\begin{equation}
D_{n^{\ast }}{}{}{}{}{}{}{}{}-\gamma \left( \frac{T-T_{C}}{T_{C}}+{}\frac{%
{}{}{}{}{}{}{}{}{}{}{}{}{}{}\left\langle n^{\ast }\right\rangle }{n}%
{}{}\right) {}\frac{{}{}{}{}{}{}{}{}{}{}{}{}{}{}\left\langle n^{\ast
}\right\rangle }{n}{}{}=0;\quad \frac{{}\partial u_{S}(R,t)}{{}\partial R}=0.
\label{12.12.24}
\end{equation}
This equation defines $\left\langle n^{\ast }\right\rangle $ at all values
of temperature. From the second equation follows, that the superconducting
current is vortical.

\section{Dissipative London equation}

With the help of the kinetic equation we shall receive the equation for
velocity:

\begin{equation}
\frac{\partial u_{S}}{\partial t}+\left( u_{S}\frac{\partial }{\partial R}%
\right) u_{S}=\nu \frac{\partial ^{2}u_{Si}}{\partial R^{2}}+\frac{e^{\ast }%
}{m^{\ast }}\left( E(R,t)+\frac{1}{c}\left[ u_{S}B(R,t)\right] \right)
;\quad divE=0  \label{12.12.26}
\end{equation}
We assume, that coefficient of diffusion, of viscosity and of temperature
conductivity are identical. This gives the basis for replacement $%
D\rightarrow \nu .$

At weak currents and weak magnetic field it is possible to neglect the
nonlinear terms and to write down the equation for hydrodynamic velocity of
electron pairs the following equation:

\begin{equation}
\frac{\partial u_{S}}{\partial t}=\nu \frac{\partial ^{2}u_{S}}{\partial
R^{2}}+\frac{e^{\ast }}{m^{\ast }}E.  \label{12.12.27}
\end{equation}
In a combination with the Maxwell equations:

\begin{equation}
rotB=\frac{4\pi }{c}j_{S},\quad rotE=-\frac{1}{c}\frac{\partial B}{\partial t%
},\quad divE=0  \label{12.12.28}
\end{equation}
we have the closed system of the equations for density and hydrodynamical
velocity of electron pairs, and also for electrical and magnetic fields.

Thus we have '' the dissipative London equation'' (\ref{12.12.27}). How to
explain presence of a superconducting current in a dissipative system?

In the following section the question on connection of an opportunity of
existence of a superconductive electric current and the presence of the
flicker noise - "of noise $1/f$. will be discussed. The assumption, that the
flicker noise defines an opportunity existence coherent state -
superconductivity, on first sight seems as paradoxical. However we shall
see, that the flicker-noise presents a coherent state similar, in some
measure, to condensation of bose-gas. Now, however, "condensation" in space
of wave numbers occurs in a dissipative medium.

\section{Flicker noise and superconductivity}

The physical phenomenon ''flicker noise'' ('' $1/\omega "$ noise) consists
in abnormal behavior of a spectrum of fluctuation of any physical
characteristics in the region of low frequencies $\omega .$ (Kogan, 1985;
Klimontovich, 1982, 1983). The flicker noise is characterized as well is by
abnormal large of time correlations $\tau _{cor}$.

In spite of on efforts of many scientists until now absent the unified point
of view on the nature of the flicker noise. Our point of view on this
problem is presented in (Klimontovich, 1982, 1983, 1995).

The flicker noise exists in the region of frequencies is restricted from the
side of large frequencies by the diffusion time $\tau _{D}=L^{2}/D.$ Here $D$
- the coefficient of spatial diffusion, and $L-$ the minimal characteristic
scale of sample.

For the minimal frequency $\omega _{\min }$ there are possible two
definitions - ''subjective '', dependent from the observation time $\tau
_{obs}$, and ''objective '', is defined by the parameters of all system.
Experience shows that the dependence $1/\omega $ conserves in during of the
increasing of the observation time. This gives the reason to suppose that
the minimal frequency of the flicker noise spectrum is defined by the time
life $\tau _{life}.$ Thus, the region of the flicker noise is defined by
inequalities:

\begin{equation}
\frac{1}{\tau _{life}}\leq \frac{1}{\tau _{obs}}\ll \omega \ll \frac{1}{\tau
_{D}}=\frac{D}{L^{2}}.  \label{12.13.3}
\end{equation}
In the flicker noise domain appears the new scale and corresponding volume:

\begin{equation}
L_{\omega }=\sqrt{\frac{D}{\omega }}\gg L,\qquad V_{\omega }=L_{\omega
}^{3}\gg V.  \label{12.13.4}
\end{equation}
At these conditions the dimension of a sample $d$ is not play the role and
in the limit of very large observation times it can count equal to zero,
that corresponds to zero volume $V$. Below under $L$ we shall use the
minimal from characteristic scales.

Thus. we have the chain of inequalities for volumes 
\begin{equation}
V=V_{D}\ll V_{\omega }\ll V_{obs}\leq V_{life}.  \label{12.13.5}
\end{equation}
According to stated representations the equilibrium flicker noise arises at
diffusion in limited volume. In the unlimited sample appropriate spectral
density at diffusion is defined by known expression:

\begin{equation}
\left( \delta n\delta n\right) _{\omega ,k}=\frac{\left( yy\right) _{\omega
,k}}{\omega ^{2}+\left( Dk^{2}\right) ^{2}},\quad \left( yy\right) _{\omega
,k}=2Dk^{2}\left( \delta n\delta n\right) _{k}.  \label{12.13.8}
\end{equation}
Designation for the spectral density of appropriate the Langevin source here
is entered. For ideal gas $\left( \delta N\delta n\right) _{k}=n.$ In a
general case it is defined through the isothermal compressibility of a
sample.

For domain of the flicker noise are carried out the following inequalities $%
\sqrt{D/\omega }\gg L,V_{\omega }\gg V.$ First of them is possible to write
down as $1/\omega \gg \tau _{D}=L^{2}/D.$ The spectral density of the
appropriate the Langevin source is defined by expression (Klimontovich,
1990abc, 1995)

{\sl 
\begin{equation}
\left( yy\right) _{\omega ,k}=2Dk^{2}AV_{\omega }\left\langle \delta
n_{V}\delta n_{V}\right\rangle \exp \left( -\frac{Dk^{2}}{2\omega }\right)
\qquad n_{eff}=AV_{\omega }\left\langle \delta n_{V}\delta
n_{V}\right\rangle .  \label{12.13.10}
\end{equation}
}Here $\left\langle \delta n_{V}\delta n_{V}\right\rangle $ - correlator of
fluctuations, averaged on volume $V$. For ideal gas $\left\langle \delta
n_{V}\delta n_{V}\right\rangle =n/V.$ 

Appropriate expression for intensity of a Langevin source has a form: 

\begin{equation}
\left( yy\right) _{\omega ,k}=2Dk^{2}An\frac{V_{\omega }}{V}\exp \left( -%
\frac{Dk^{2}}{2\omega }\right) .  \label{12.13.13}
\end{equation}
Thus, for the domain of flicker of noise has place the replacement 

\begin{equation}
n\rightarrow n_{eff}=An\frac{V\omega }{V}.  \label{12.13.14}
\end{equation}
By this the repeated diffusion is taken into account - each particle covers
the diffusion volume $V_{\omega },$ which is a lot of greater volume of a
sample $V$. 

Thus each particle ''works'' much times. It also is resulted that the
effective density of particles in $V_{\omega }/V$ times more of real one.
This is taken into account by replacement (\ref{12.13.14}). The constant
multiplier $A$ will be determined from a normalization condition. 

In intensity of a Langevin source there is a strong dependence from
frequencies and stronger dependence on wave number. Thus the dispersion on
wave numbers is proportional to frequency $\omega $: 

\begin{equation}
\left\langle \left( \delta k\right) ^{2}\right\rangle \sim \frac{1}{%
L_{\omega }^{2}}=\frac{\omega }{D},  \label{12.13.15}
\end{equation}
therefore in a domain of flicker noise arises a very sharp distribution on
to wave numbers - original Bose-condensation. It speaks about that in the
field of the flicker noise arises the spatial coherence. 

Substitution of expression (\ref{12.13.14}) in the first formula ( \ref
{12.13.8}) results in expression for spatially temporary spectral density in
the region of flicker noise: 

\begin{equation}
\left( \delta n\delta n\right) _{\omega ,k}=\frac{2Dk^{2}}{\omega
^{2}+\left( Dk^{2}\right) ^{2}}AV_{\omega }\left\langle \delta n_{V}\delta
n_{V}\right\rangle \exp \left( -\frac{Dk^{2}}{2\omega }\right) .
\label{12.13.16}
\end{equation}
Here it is possible to execute integration on $k$ and to receive expression
for appropriate temporary spectral density: 

\begin{equation}
\left( \delta n\delta n\right) _{\omega }=\frac{\pi \left\langle \delta
n_{V}\delta n_{V}\right\rangle }{\ln \left( \tau _{life}/\tau _{D}\right) }{}%
\frac{1}{\omega },\qquad \frac{1}{\tau _{life}}\leq \frac{1}{\tau _{obs}}\ll
\omega \ll \frac{1}{\tau _{D}}.  \label{12.13.17}
\end{equation}
The constant multiplier $A$ is determined from a the normalization
condition: 
\begin{equation}
\int_{1/\tau _{life}}^{1/\tau _{D}}\left( \delta n\delta n\right) _{\omega }%
\frac{d\omega }{\pi }=\left\langle \delta n_{V}\delta n_{V}\right\rangle .
\label{12.13.18}
\end{equation}
It is supposed, thus, that the basic contribution to the correlator $%
\left\langle \delta n_{V}\delta n_{V}\right\rangle $ is in the domain of the
flicker noise. 

\subsection{Residual temporary correlation}

The temporary correlation is connected to temporary spectral density by
relation: 

{\sl 
\begin{equation}
\left\langle \delta n\delta n\right\rangle _{\tau }=\int_{1/\tau
_{life}}^{1/\tau _{D}}\left( \delta n\delta n\right) _{\omega }\frac{d\omega 
}{\pi }.  \label{12.13.19}
\end{equation}
}

From here follows, that 
\[
\left\langle \delta n\delta n\right\rangle _{\tau }=\left( C-\frac{\ln
\left( \tau /\tau _{D}\right) }{\ln \left( \tau _{life}/\tau _{D}\right) }%
\right) \left\langle \delta n_{V}\delta n_{V}\right\rangle \text{ at }\tau
_{D}\ll \tau \ll \tau _{life},
\]
\begin{equation}
\label{12.13.20}
\end{equation}
\[
C=1-\frac{\gamma }{\ln \left( \tau _{life}/\tau _{D}\right) },\quad \text{ }%
\gamma =0.577.
\]
Here are used the Euler constant. 

Thus, in the domain of the flicker noise dependence from $\tau $ very much
weak - logarithmic at the large value of argument. It gives the basis to
speak about presence of residual temporary correlations. 

Let's define the appropriate time of correlation: 

{\sl 
\begin{equation}
\tau _{cor}=\int_{\tau _{D}}^{\tau _{life}}\left\langle \delta n\delta
n\right\rangle _{\tau }d\tau /\left\langle \delta n\delta n\right\rangle
_{\tau =\tau _{D}}.  \label{12.13.21}
\end{equation}
}

For an estimation of correlation time is spent the integration at
performance of the inequalities $\tau _{D}\ll \tau \ll \tau _{life}.$ In
result we find, that 

\begin{equation}
\tau _{cor}\sim \tau _{life}/\ln \frac{\tau _{life}}{\tau _{D}}.
\label{12.13.23}
\end{equation}
Thus, the time of correlation at the unlimited time life $\tau _{life}$
aspires to infinity. 

Stated gives the basis for a conclusion, that in the region of the flicker
noise has a place as spatial, and temporary coherence. It also opens an
opportunity for an establishment connection of two coherent phenomena: the
flicker noise and the superconductivity. 

\subsection{Flicker noise and superconductivity}

The flicker noise represents spatial - temporary coherent structure. It
makes existence of connection between the flicker noise and the
superconductivity less surprising. 

In approximation of the first moments from the equation (12.12.10) was
received the appropriate system of the hydrodynamical equations. It consists
from two equations of a continuity with '' a chemical source '' (12.12.23),
which at condition $\left\langle n_{S}\right\rangle _{R,t}=const.$ is
equivalent to the equations (\ref{12.12.24}). First of them allows to find
dependence of this quantity $\left\langle N_{S}\right\rangle _{R,t}$ from
temperature. Second shows, that the velocity field is vortical. 

The equation for average velocity we shall rewrite as the equation for a
whirlwind of electrical current and also we shall enter into it the
appropriate Langevin source: 

\begin{equation}
\frac{\partial }{\partial t}\left[ \Omega +\frac{e^{\ast 2}n_{S}}{m^{\ast }c}%
B\right] =\nu \frac{\partial ^{2}\Omega }{\partial R^{2}}+y_{\Omega
}(R,t),\quad \Omega =rotj,\quad D=\nu .  \label{12.13.24}
\end{equation}
This equation should be complemented by the Maxwell equation: 
\begin{equation}
rotB=\frac{4\pi }{c}j.  \label{12.13.25}
\end{equation}

Thus for the whirlwind we have the equation of diffusion type with the
Langevin source?. For the spectral density of the Langevin source it is
passible to use the expression similar to formula (\ref{12.13.10}) 

{\sl 
\begin{equation}
\left( y_{\Omega }y_{\Omega }\right) _{\omega ,k}=2\nu k^{2}AV\omega \exp
\left( -\frac{\nu k^{2}}{2\omega }\right) \left\langle \delta \Omega
_{V}\delta \Omega _{V}\right\rangle .  \label{12.13.26}
\end{equation}
}

The normalization condition has the form: 

{\sl 
\begin{equation}
\int_{1/\tau _{life}}^{1/\tau _{D}}\frac{d\omega }{\pi }\int \frac{dk}{(2\pi
)^{3}}\left( \delta \Omega \delta \Omega \right) _{\omega ,k}=\left\langle
\delta \Omega _{V}\delta \Omega _{V}\right\rangle .  \label{12.13.27}
\end{equation}
}

Now we need the equation for the description of temporary evolution of a
whirlwind of average current. 

Let's return to the formula (\ref{12.13.26}) \} and we shall rewrite it as
the fluctuation dissipation relation (FDR): 

\begin{equation}
\left( y_{\Omega }y_{\Omega }\right) _{\omega ,k}=2\gamma (\omega
,k)A\left\langle \delta \Omega _{V}\delta \Omega _{V}\right\rangle .
\label{12.13.28}
\end{equation}
The designation for appropriate dissipative coefficient here is entered at
presence both temporary, and spatial dispersion 

\begin{equation}
\gamma (\omega ,k)=\nu k^{2}V\omega \exp \left( -\frac{\nu k^{2}}{%
{}{}{}2\omega }\right) .  \label{12.13.29}
\end{equation}
By return the Fourier transformation it is possible to receive expression
for appropriate the dissipative operator. We use its elementary
representation in form $"1/\tau _{rel}"$ . 

In this approximation the relaxation time about time of correlation, i.e. 

\begin{equation}
\tau _{rel}\sim \tau _{cor}\sim \tau _{life}/\ln \frac{\tau _{life}}{\tau
_{D}}\gg \tau _{obs}\gg \tau _{D}.  \label{2.12.30}
\end{equation}

In result for a whirlwind of an electrical current instead of (12.13.24) is
received the following ''model'' equation for average value of a whirlwind 

\begin{equation}
\frac{\partial }{\partial t}\left[ \Omega +\frac{e^{\ast 2}n_{S}}{m^{\ast }c}%
B\right] =-\frac{1}{\tau _{rel}}\Omega ,\quad \Omega =rotj.  \label{12.13.32}
\end{equation}
It should be solved together with the Maxwell equation (\ref{12.13.25}). 

As the relaxation time is the order of the time life of installation, in
zero approximation on dimensionless parameter 

\begin{equation}
\frac{\tau _{obs}}{\tau _{life}}\quad \text{at \quad }\tau _{life}\gg \tau
_{obs}\gg \tau _{D}  \label{12.13.33}
\end{equation}
in the equation (\ref{12.13.32}) it is possible to neglect by dissipation.
We come, such to the equation 
\[
\frac{\partial }{\partial t}\left[ \Omega +\frac{e^{\ast 2}n_{S}}{m^{\ast }c}%
B\right] =0.
\]

At integration on time a constant of integration it is possible to accept
for zero, that will be coordinated to the Meissner. In result we come to the
London equation 

\begin{equation}
rotj=\frac{e^{2}n_{s}}{mc}B.  \label{12.13.35}
\end{equation}
We consider it together with the Maxwell equation (\ref{12.13.25}). It
enables to receive the closed equation for a magnetic field 

\begin{equation}
\frac{\partial ^{2}B}{\partial R^{2}}-\frac{1}{\delta _{L}^{2}}B=0,
\label{12.13.36}
\end{equation}
which, as we already know, describes the Meissner effect. 

So, many times diffusion and, as a consequence,  the flicker noise allows to
understand an opportunity of simultaneous existence of the not fading
electrical current and screening of a magnetic field. Here it is difficult
to tell, which of these two phenomena is more fundamental, so they are bound
among themselves. Both of them appear possible thanking spatially temporary
coherent fluctuations of a whirlwind of a current or magnetic field. 

Let's return to inequalities (\ref{12.13.33}). We shall estimate values of
the observation tines, at which becomes possible to observe ''not fading''
superconducting electrical current. 

The diffusion time $\ \tau _{D}$ \ defines $\omega _{\max }$ \ for domain of
the flicker noise: $\omega _{\max }\sim 1/\tau _{D}.$ The minimal scale is
the London length $\delta _{L}\sim $ 10 $^{\text{-5}}$ cm. The diffusion
coefficient can be estimated by one of two formulas: $D\approx \hbar
/2m^{\ast };$ $v_{F}l$ ($l$ - the effective length of free paths of electron
pairs). From these formulas follows, that 
\begin{equation}
\left( \tau _{obs}\right) _{\min }\geq D=\delta _{L}^{2}/D\sim \text{10}^{%
\text{-10}}\text{-10}^{\text{-11}}\sec .  \label{12.13.37}
\end{equation}

Let's address now to definition of concepts ''not fading'' or
''superconductive '' current. 

One of definition of these concepts is ''measuring''. It is connected with
the observation time. Outside of the observation time it is impossible to
guarantee the constancy of a current. 

However, as in process of increase of the observation time it is impossible
to find out attenuation of a current, it is natural to assume, that the
constancy of a current has a place within the limits of greatest temporary
interval $\tau _{life},$ i.e. '' the time life '' of installations. 

\section{Conclusion}

So, the attempt is made to connect two, apparently, incompatible phenomena:
the existence of the flicker noise and the superconductivity in a
dissipative medium. 

The flicker noise arises due to formation of the spatial coherent structure
at the many times diffusion process. Due to this the dissipation, caused
with viscous friction, is replaced on dissipation with characteristic time
about time of life of installation. It also opens an opportunity for
existence not fading (within the limits of time life of installation)
current, and also of the screening of a magnetic field. 

The connection of the flicker noise and superconductivity allows to spill
additional light on the question on an opportunity of existence
superconducting current in the dissipative medium. According to stated, the
occurrence of the flicker-noise, and consequently also superconductivity, is
promoted by the smallest scale of length. 

In the theory, considered above, the role of such parameter play the London
length. In this connection it is possible to expect, that preferable to
occurrence of superconductivity are the substances having layered structure. 

At last, the phase transition  results to occurrence of quantum continuous
medium, for which the basic small parameter is the quantum the physical
Knudsen number. 

{\bf REFERENCES }

1.Abrikosov A.A.., Foundation of theory of metals (''Nauka'' Moscow 1987 (in
Russian)). 

2 .Bardeen J., Cooper L., Schriffer J. Theory of superconductivity
(Phys.Rev. 106 (1957) 162). 

3. Bednorz J., Mueller K. (The Nobel Foundation, 1988). 

4. Bogolubov N.N. (JETP 34 (1958) 58; JETF 34 (1958) 71). 

5. Elesin V.F., Kopaev Yu.V. Uspechi Fiz. Nauk 133 (1981) 259). 

6. Feinman R. The Feinman Lectures on Physics. V.9 (Addison, Massachusetts,
London, 1963). 

7. Ginsburg V.L. Landau L.D. (JETP, 20 (1950) 1064). 

8. Gor'kov L.P. (JETP 34 (1959) 1818.) 

9. Guliyan A.M., Jarkov G.F. Superconductors in external field (''Nauka''
Moscow, 1990). 

10. Klimontovich Yu.L. Statistical Physics (''Nauka'' , Moscow, 1982;
Harwood, New York, 1986). 

11. Klimontovich Y.L. (Pis'ma v JTP 9 (1983) 406; Sov. Techn. Phys. Lett. 9
(1983) 174). 

12. Klimontovich Y.L. (Pis'ma v JETP 51 (1990) 43). 

13. Klimontovich Yu.L. (Physica A 167 (1990) 782). 

14. Klimontovich Yu.L. Turbulent Motion and the Structure of haos
(''Nauka'', Moscow, 1990; Kluwer, Dordrecht, 1991). 

15. Klimontovich Yu.L. Statistical Theory of Open Systems. (V.I ''Yanus''
Moscow, 1995; Kluwer, Dordrecht, 1995); V.II ''Yanus'', Moscow, 1999; V.III
''Yanus'', Moscow, 2001). 

16. Kogan Sh.M. (Uspehi Fiz. Nauk 145 (1985) 286). 

17. Lifshitz E.M., Pitaevskii L.P. Statistical Physics. Part 2 (''Nauka?''
Moscow, 1978). 

18 London F. Superfluids. Vol.I Superconductivity (Wiley New York, 1950). 

20. Oraevskii A.N. (PETP 103 (1993) 262). 

21. Plakida N.M. Hightemperature superconductors (''International Program of
Education'' Moscow,1996 (in Russian)). 

22. Tinkham M. Introduction in Superconductivity (''Atomizdat'', Moscow,
1980 (in Russian)). 

\end{document}